\documentclass[conference]{IEEEtran}
\usepackage{amsmath,graphicx}
\usepackage{amsfonts,amssymb}
\usepackage{multirow}
\usepackage{algorithm}
\usepackage{algpseudocode}
\usepackage{cite}
\usepackage{balance}
\DeclareGraphicsExtensions{.jpg,pdf,.png}
\ifCLASSINFOpdf
\else
\fi
\usepackage{fancyhdr}
\newcommand*{\myfont}{\fontfamily{phv}\selectfont}
\DeclareTextFontCommand{\textmyfont}{\myfont}

\begin{document}

\title{Comparison of Uniform and Random Sampling\\for Speech and Music Signals}
\author{Nematollah Zarmehi, Sina Shahsavari, and Farokh Marvasti\\Advanced Communication Research Institute\\Department of Electrical Engineering\\Sharif University of Technology, Tehran, Iran\\Email: http://zarmehi.ir/contact.html}

\maketitle
\thispagestyle{fancy}
\begin{abstract}
In this paper, we will provide a comparison between uniform and random sampling for speech and music signals. There are various sampling and recovery methods for audio signals. Here, we only investigate uniform and random schemes for sampling and basic low-pass filtering and iterative method with adaptive thresholding for recovery. The simulation results indicate that uniform sampling with cubic spline interpolation outperforms other sampling and recovery methods.
\end{abstract}
\section{Introduction}\label{sec:intro}
Various sampling and recovery methods proposed in the field of signal processing. The Nyquist-Shannon theorem proposes condition to recover a band-limited signal from its samples \cite{ref:shann,ref:bookmarvasti,ref:booklanda}. The uniform sampling using an anti-aliasing filter and low-pass (LP) filtering were used for some decades. After that, other sampling methods such as non-uniform sampling \cite{non1,mbm,non2}, periodic non-uniform sampling \cite{pnon1,pnon2}, and random sampling \cite{rand1,rand2} were proposed.

In this paper, we provide a comparison between uniform and random sampling for speech and music signals. We use basic LP filtering and spline interpolation for uniform sampling and Iterative Method with Adaptive Thresholding (IMAT) for random sampling. 

\section{Sampling and Recovery Methods}\label{sec:algs}
This Section introduces the sampling and recovery methods that we are going to compare. We compare uniform and random sampling schemes for speech and music signals along with different recovery methods such as basic LP filtering, spline interpolation, and IMAT \cite{ref:imat,ref:imatsite}. IMATI is a version of IMAT algorithm that uses interpolation operator in each iteration \cite{ref:imati}. Table \ref{tab:schems} shows the sampling and recovery schemes used in this paper. The abbreviations AF stands for Anti-aliasing Filter.

\begin{table}[h!]
	\renewcommand{\arraystretch}{1.8}
	\centering
	\caption{Sampling and recovery schemes.}\label{tab:schems}
	\begin{tabular}{|c|c|c|}
		\hline \hline
		\bf Short Name & \bf Sampling & \bf Description \\ \hline \hline
		\bf U-AF-FFT-Sp  & AF \& Uniform & FFT as AF \& Spline interpolation \\ \hline
		\bf U-AF-FFT  & AF \& Uniform & FFT as AF \& LP-filtering \\ \hline
		\bf U-AF-FIR-Sp & AF \& Uniform & FIR as AF \& Spline interpolation \\ \hline
		\bf U-AF-FIR & AF \& Uniform & FIR as AF \& LP-filtering \\ \hline
		\bf R-IMATI   & Random & IMATI \\ \hline
		\bf R-Sp  & Random & Spline interpolation\\ \hline \hline
	\end{tabular}
\end{table}

We have used some objective performance metric criteria for comparison of above methods. They are listed in Table \ref{tab:metrs}. The recovered {\it ``.WAV"} files are also saved on memory disk for subjective evaluations.

\begin{table}[h!]{\scriptsize{
	\renewcommand{\arraystretch}{1.8}
	\centering
	\caption{Objective performance metric criteri}\label{tab:metrs}
	\begin{tabular}{|c|c|c|}
		\hline \hline
		\bf Name & \bf Quantity & \bf More description \\ \hline \hline
		\bf SNR & $dB$ & $SNR(x,\hat x)=20 \log \left({\frac{\|x\|}{\|x-\hat x\|}}\right)$ \\ \hline
		\bf PESQ & Dimensionless & A score between 1.0 (worst) up to 4.5 (best) \\ \hline \hline
		\bf CPU Time & $second$ &  Using {\it tic} and {\it toc} commands in MATLAB \\ \hline \hline
	\end{tabular}}}
\end{table}

\section{Simulation Results}\label{sec:sim}
In this section, we present simulation results. We have used 44.1--48kHz speech and music {\it{``.WAV''}} signals. Frame size is 1024 and the simulations are done in MATLAB R2015a on Intel(R) Core(TM) i7-5960X @ 3GHz with 23GB-RAM. 


\subsection{Speech Signal}
\begin{figure}[t!]
	\centering
	\includegraphics[width=1\linewidth]{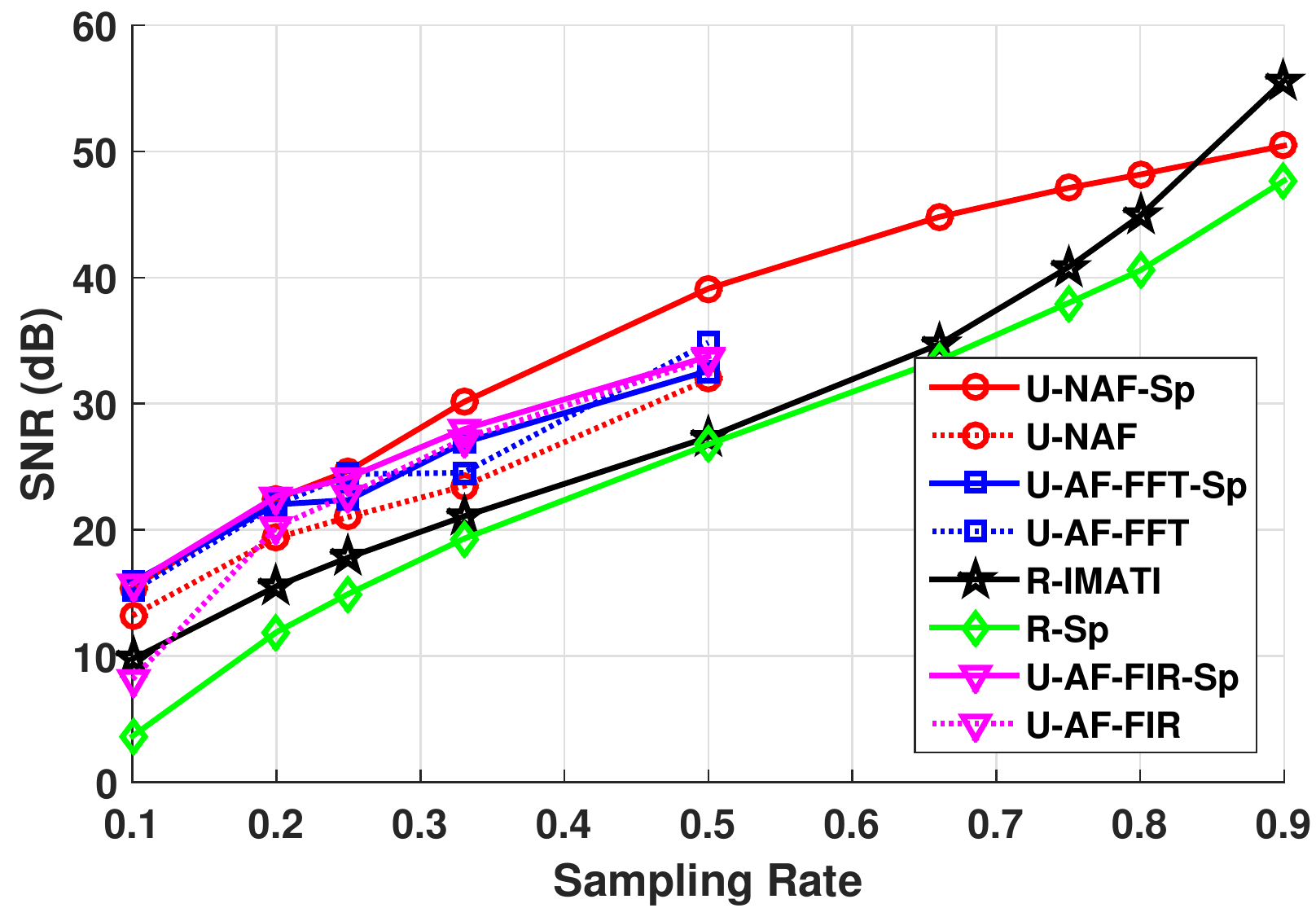}
	\caption{SNR vs. sampling rate.}
	\label{fig:snr}
\end{figure}

All sampling and recovery methods are simulated on our speech dataset and the results are presented in Fig. \ref{fig:snr} in terms of SNR. 

Fig. \ref{fig:snr} shows the SNR of all methods vs. sampling rate. In uniform sampling scheme, we used periodic uniform sampling for sampling rates greater than 0.5. According to Fig. \ref{fig:snr}, uniform sampling with spline interpolation outperforms the other methods. Another observation is that spline interpolation works well with uniform samples but its performance degraded in case of random sampling.

\begin{figure}[t!]
	\centering
	\includegraphics[width=1\linewidth]{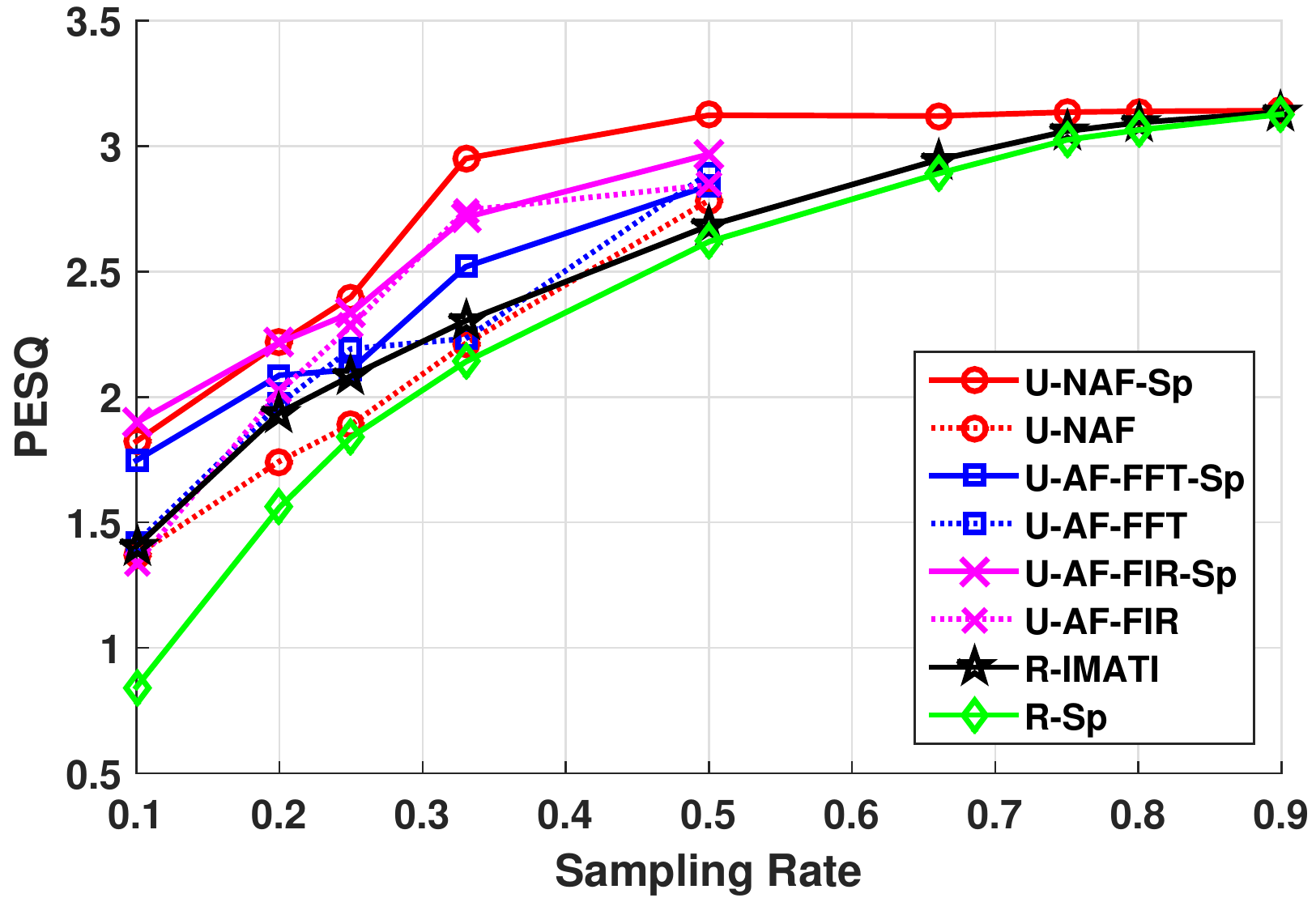}
	\caption{PESQ vs. sampling rate.}
	\label{fig:pesq}
\end{figure}

The Perceptual Evaluation of Speech Quality (PESQ) metric is employed to assess the quality of recovered speech signals. PESQ is approved as ITU-T Rec. P.862 \cite{ref:pesq}. The voice quality is rated by a value ranging from 1 (bad) to 5 (excellent). The results are shown in Fig \ref{fig:pesq}.

\subsection{Music Signal}
\begin{figure}[t!]
	\centering
	\includegraphics[width=1\linewidth]{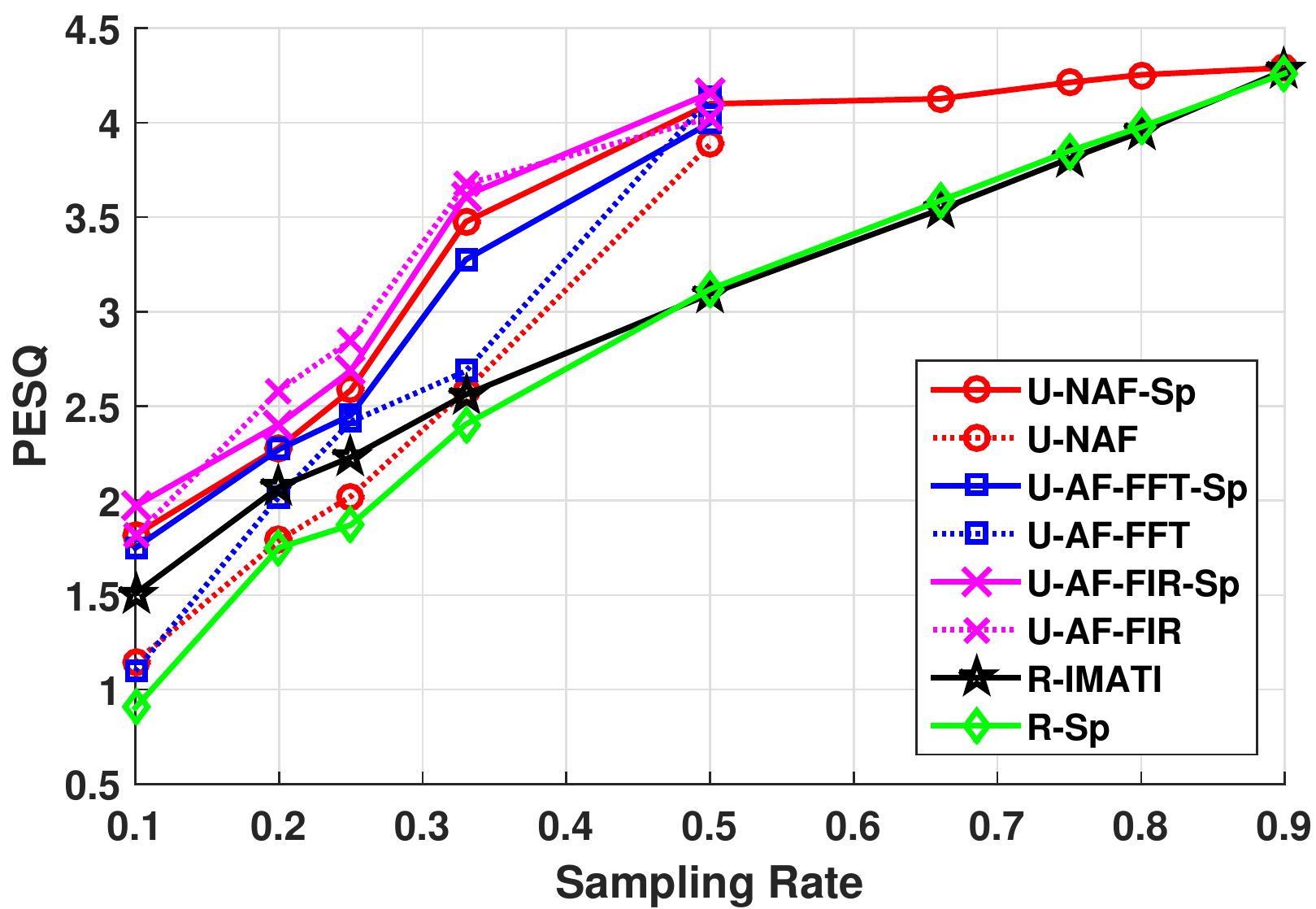}
	\caption{PESQ of all methods.}
	\label{fig:pesqm}
\end{figure}

We have also compared uniform and random sampling with music signals. The SNR of all methods is compared in Fig. \ref{fig:pesqm}. In uniform sampling scheme, the high frequency components will be filtered; Therefore, we expect spline does not work for music signal as well as for speech signal.

\begin{figure}[t!]
	\centering
	\includegraphics[width=1\linewidth]{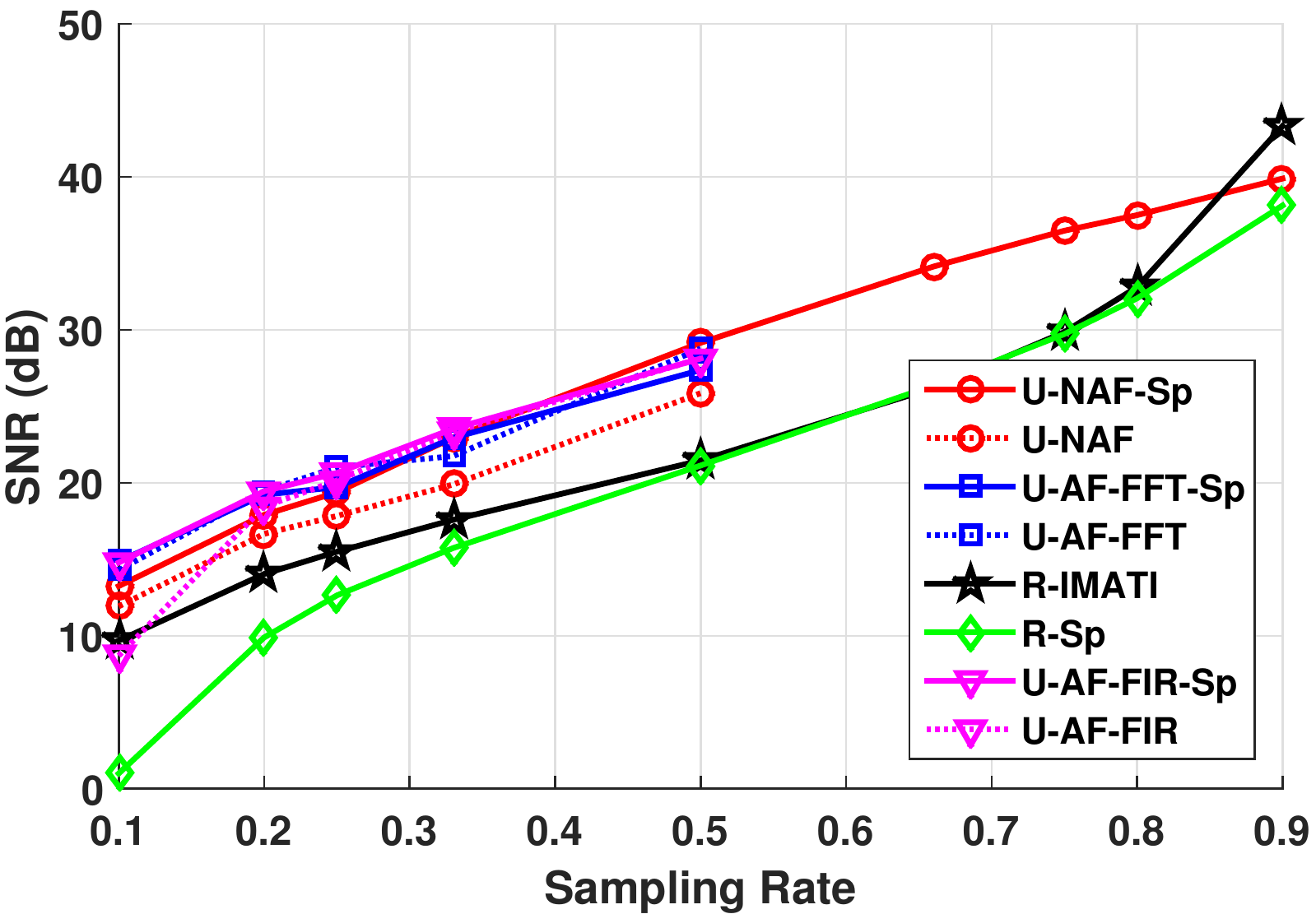}
	\caption{SNR of all methods.}
	\label{fig:snrm}
\end{figure}

Although the PESQ is not coincident with music, we measured PESQ for the recovered music signals. Fig. \ref{fig:pesqm} presents the PESQ of all methods. It can be seen that uniform sampling with FIR filter as anti-aliasing filter has the best value of PESQ between these sampling and recovery methods.

%
%
%

\begin{figure}[t!]
	\centering
	\includegraphics[width=1\linewidth]{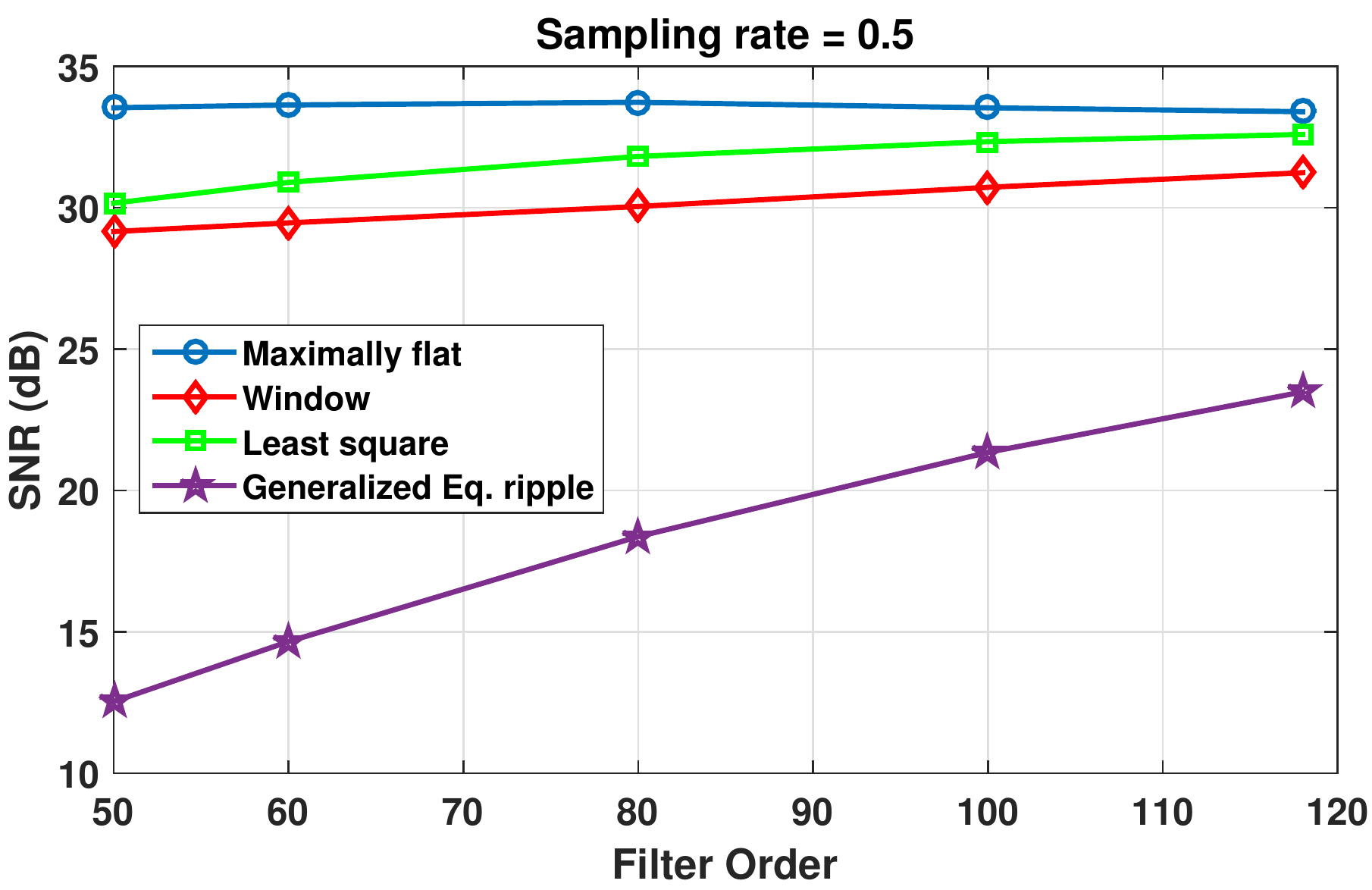}
	\caption{Different FIR Filters.}
	\label{fig:snrfilter}
\end{figure}

\begin{figure}[t!]
	\centering
	\includegraphics[width=1\linewidth]{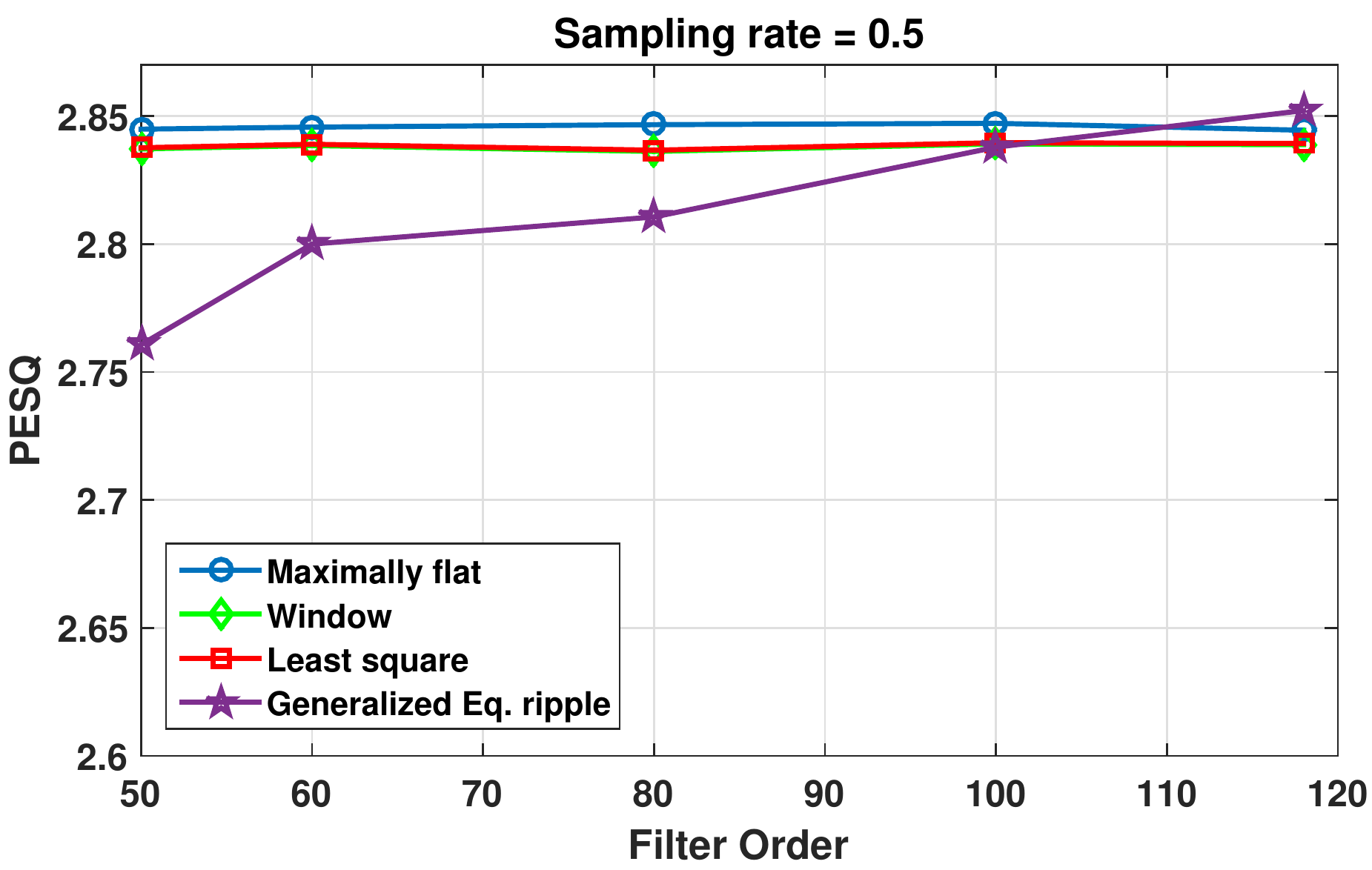}
	\caption{Different FIR Filters.}
	\label{fig:pesqm1}
\end{figure}

\begin{figure}[t!]
	\centering
	\includegraphics[width=1\linewidth]{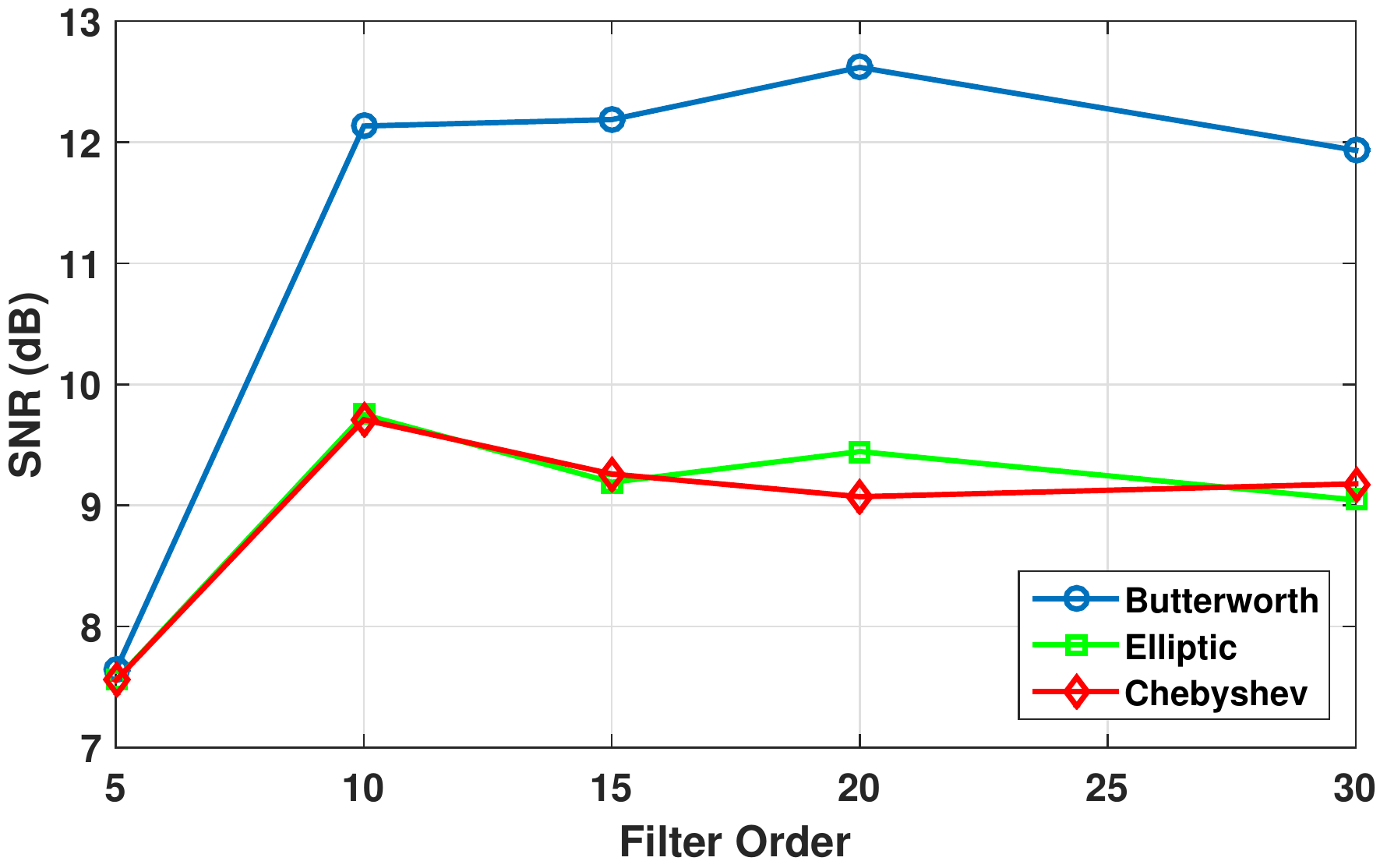}
	\caption{Different IIR Filters.}
	\label{fig:pesqm2}
\end{figure}

\begin{figure}[t!]
	\centering
	\includegraphics[width=1\linewidth]{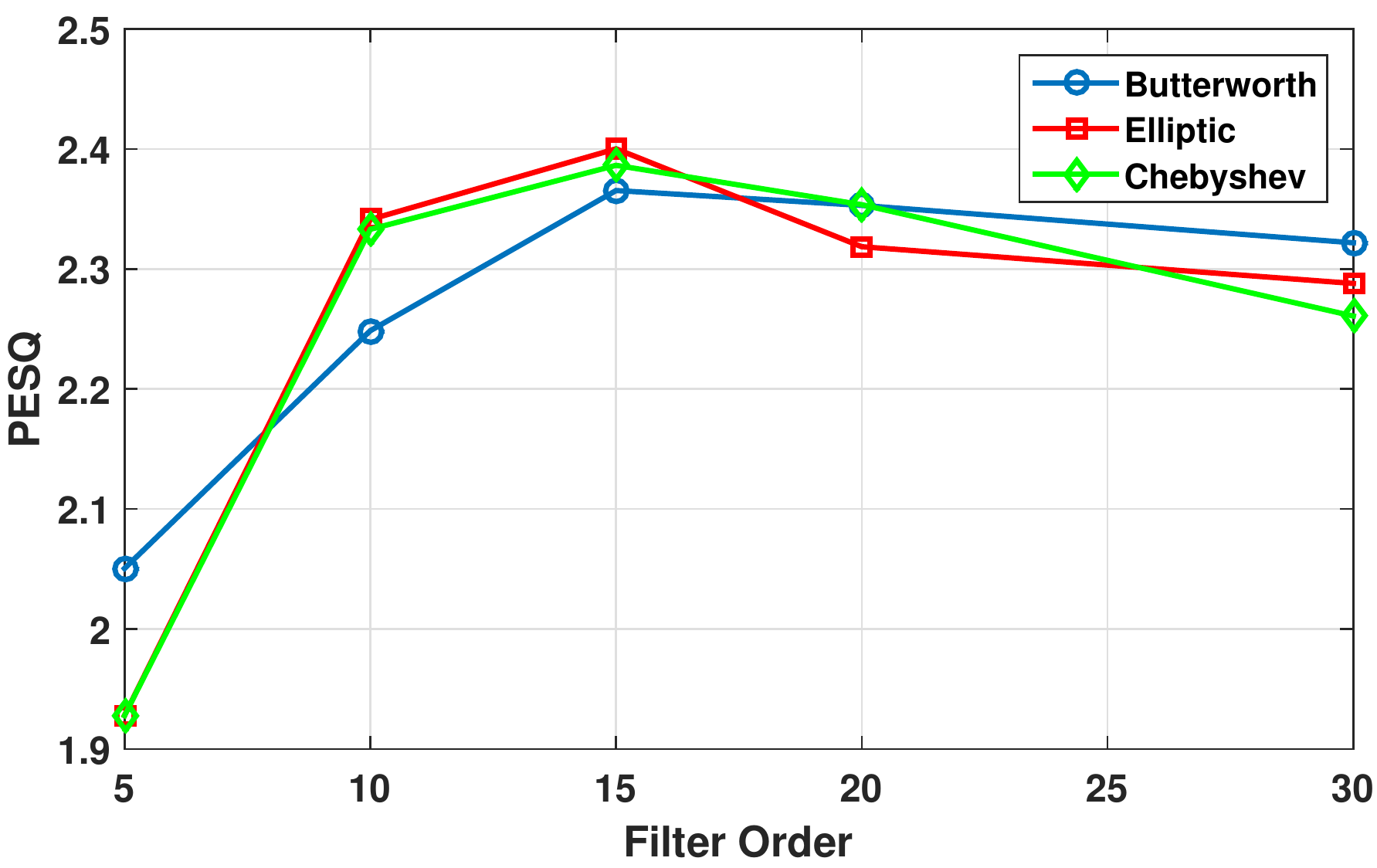}
	\caption{Different IIR Filters.}
	\label{fig:pesqfilter}
\end{figure}

We also test different FIR and IIR filters as anti-aliasing filter before uniform sampling. The original signal is recovered by cubic spline interpolation. The results are shown in Figs. \ref{fig:snrfilter}-\ref{fig:pesqfilter}. It can be seen that using FIR filter as anti-aliasing filter leads to better performance in terms of SNR and PESQ.

\subsection{Complexity Comparison}\label{subsec:pesq}
\begin{figure}[t!]
	\centering
	\includegraphics[width=1\linewidth]{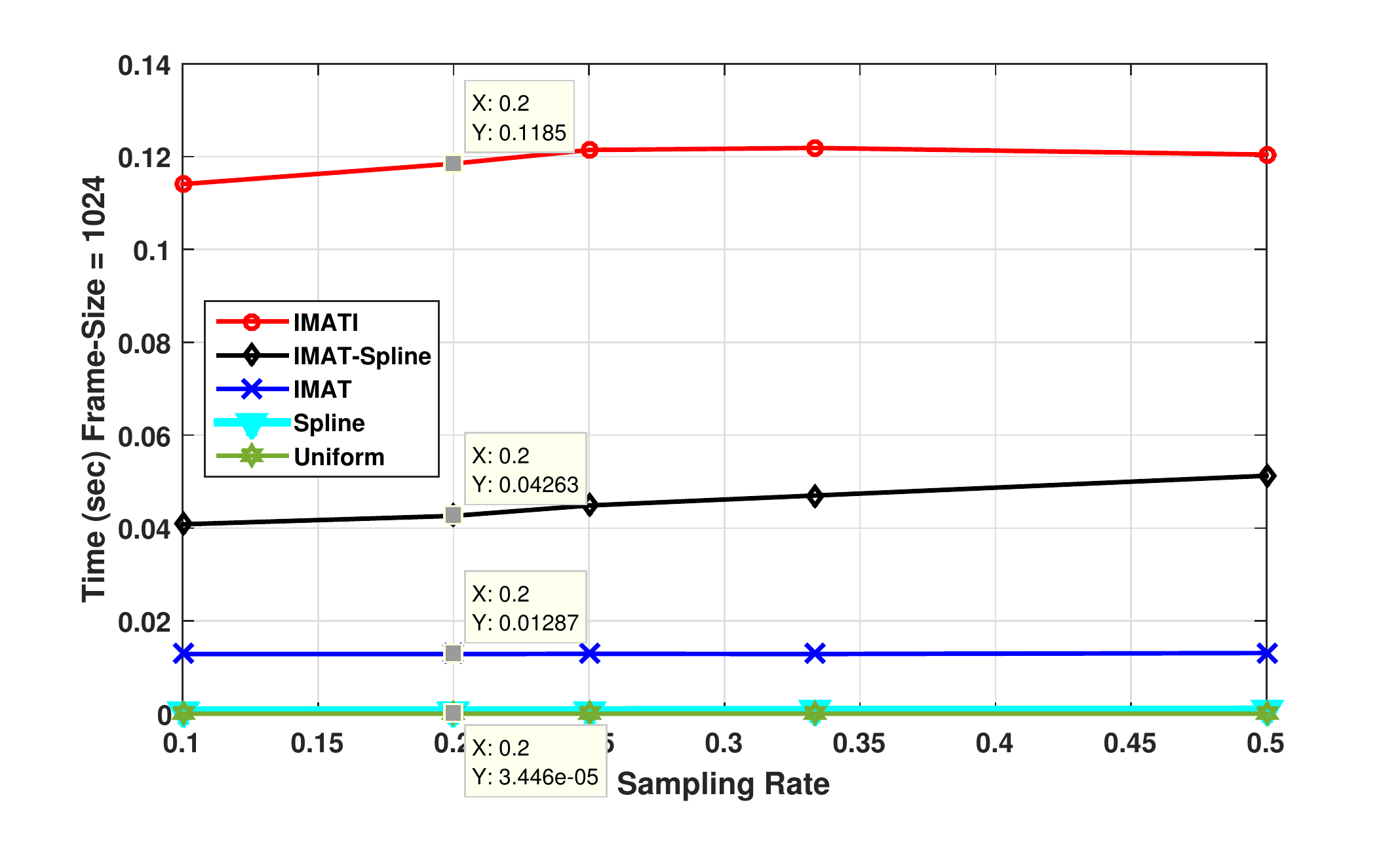}
	\caption{CPU time of all methods.}
	\label{fig:time}
\end{figure}

We provide a complexity comparison between the methods of Table \ref{tab:schems} by measuring the run time of simulations. Fig. \ref{fig:time} shows the run time for different sampling rate. It can be seen that the simulation time does not change as sampling rate changes. This is true almost for all methods.

\begin{figure}[t!]
	\centering
	\includegraphics[width=1\linewidth]{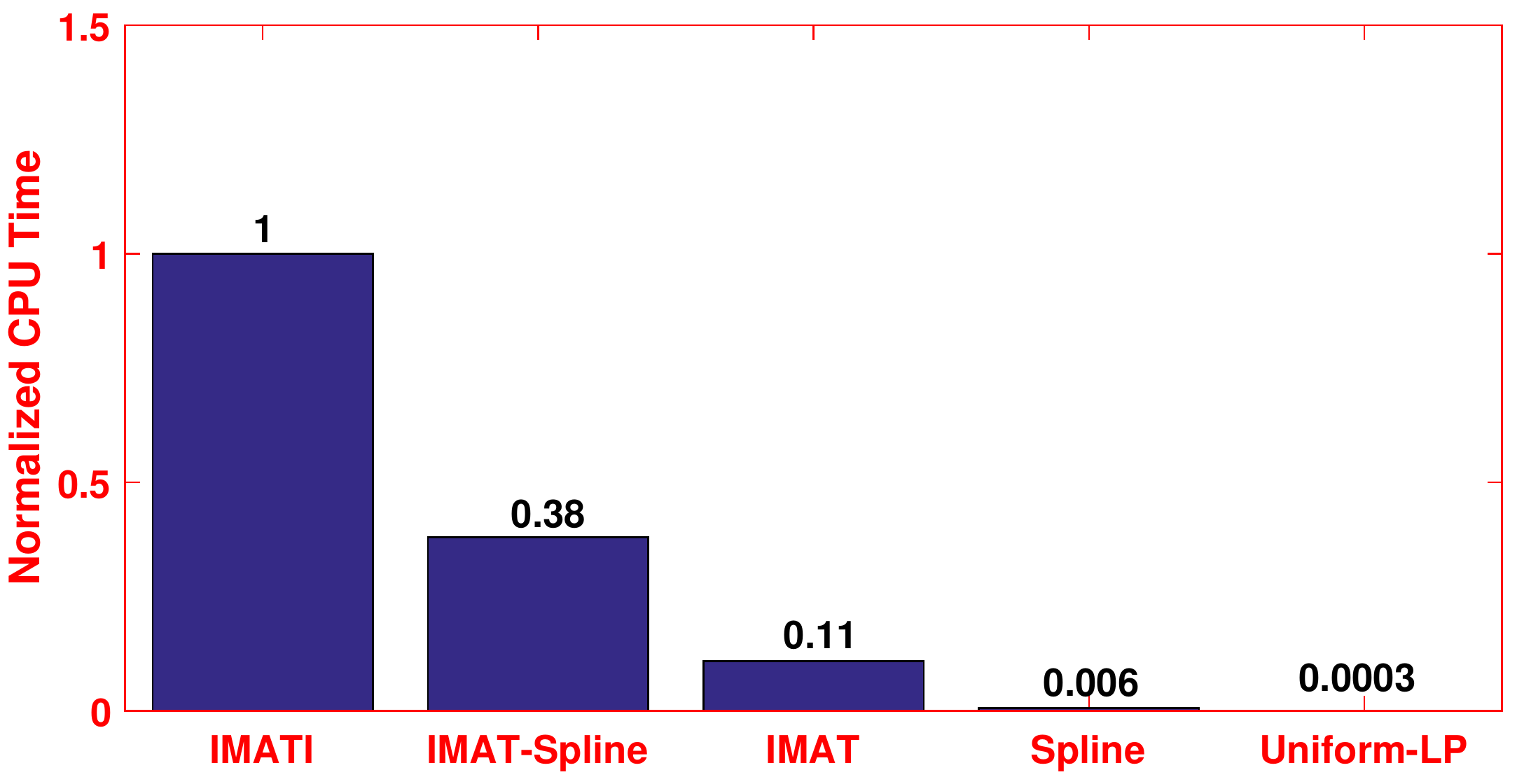}
	\caption{Normalized CPU time of all methods.}
	\label{fig:timen}
\end{figure}

The normalized CPU time is also charted in Fig. \ref{fig:timen}. Random sampling with IMATI for reconstruction takes more time than the uniform sampling with basic LP filtering or spline interpolation.

\section{Cubic Spline Interpolation}
\begin{figure}[h!]
	\centering
	\includegraphics[width=1\linewidth]{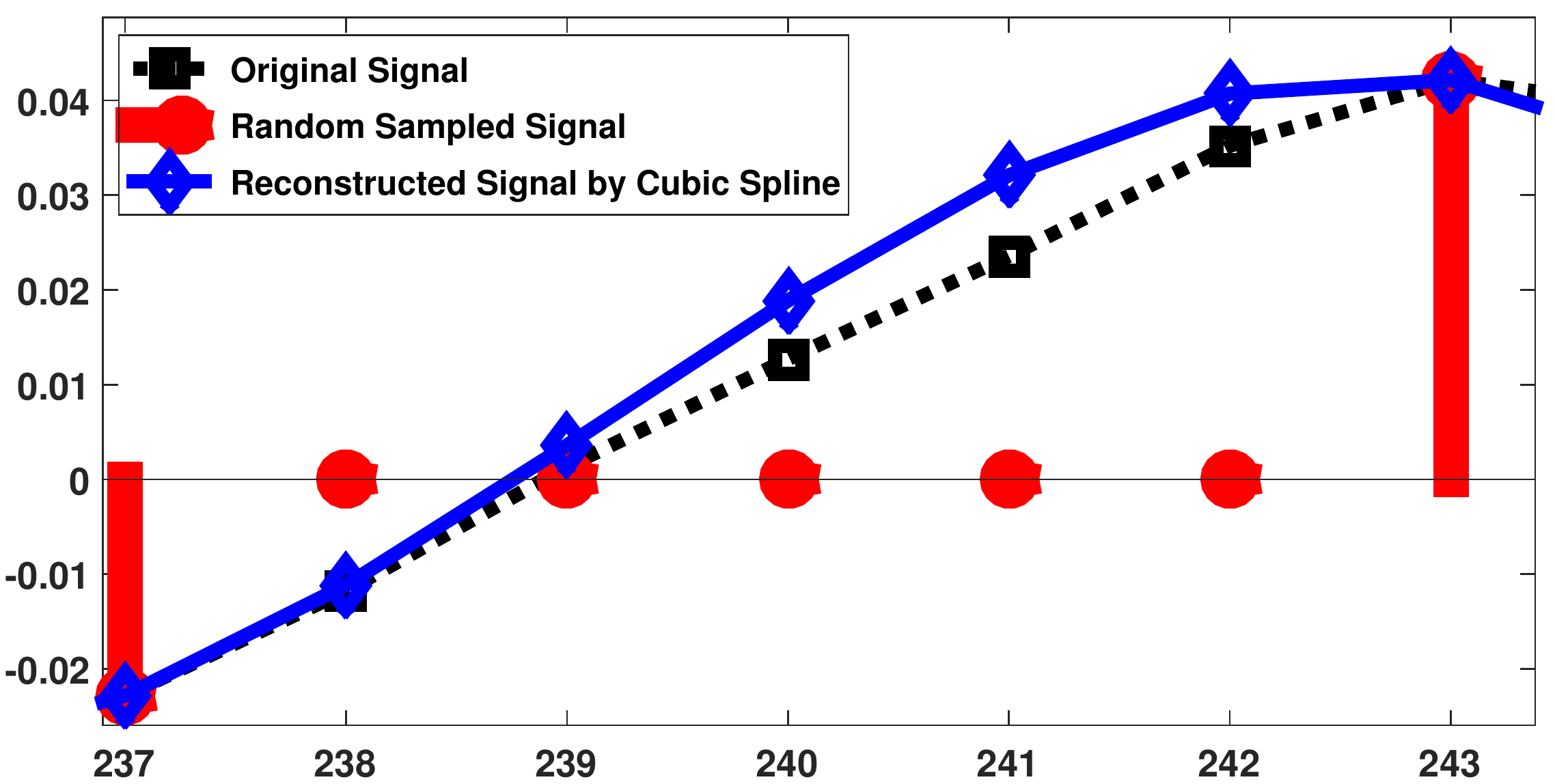}
	\caption{An example of cubic spline interpolation.}
	\label{fig:howsp}
\end{figure}

We observed that the spline interpolation does not work well in random sampling case. Among all spline interpolations, the cubic spline gives smoother interpolating polynomial that well matches to the samples. Thus, in cases that the signal is band-limited, the cubic spline leads to smaller error. As the name suggests, in the cubic spline, the interpolated value at each missing sample is a cubic interpolation of the values at neighboring samples. Given a set of $n+1$ data points $(x_i,y_i)$ where $x_0<x_1<\ldots<x_n$, the spline interpolator $S(x)$ is a polynomial of degree 3 on each subinterval $[x_{i-1},x_i]$ where $i=1,\ldots,n$, i.e.,
\begin{equation}
S(x) = \left\{ {\begin{array}{*{20}{c}}
	\begin{array}{l}
		{C_1}(x)\\
		\\
		{C_i}(x)\\
		\\
		{C_n}(x)
		\end{array}&\begin{array}{l}
		{x_0} \le x \le {x_1}\\
		\\
		{x_{i - 1}} \le x \le {x_i}\\
		\\
		{x_{n - 1}} \le x \le {x_n},
		\end{array}
		\end{array}} \right.
\end{equation}
where $C_i(x)=a_i+b_ix+c_ix^2+d_ix^3~ (d_i \neq 0)$ is a cubic function. An example of cubic spline interpolation is shown in Fig. \ref{fig:howsp}. 

%
%

\begin{figure}[h!]
	\centering
	\includegraphics[width=1\linewidth]{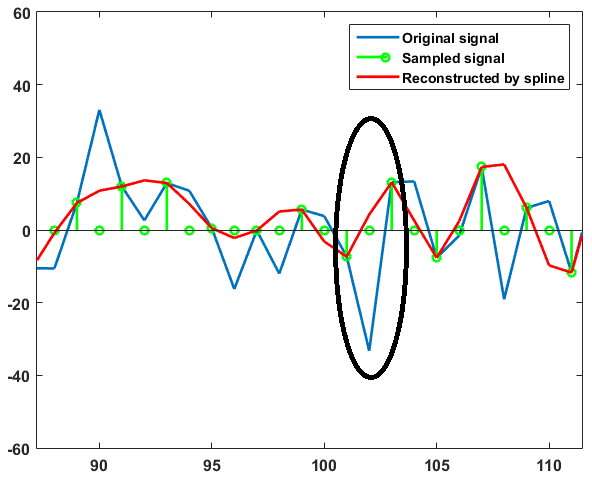}
	\caption{An example of cubic spline interpolation for pure sparse signal.}
	\label{fig:splineForSparse}
\end{figure}

We have generated an artificial 64-sparse signal with length 1024 and sampled its inverse transformed version uniformly. The original, sampled, and reconstructed signal is shown in Fig. \ref{fig:splineForSparse}. It can be seen that the original signal is not smooth at all. In other words, it is not a nearly band-limited or LP signal. Consider two samples highlighted in Fig. \ref{fig:splineForSparse}. We expect the original signal have a value near zero that is between the value of these two samples. But it is about -30.

%
%
%

\section{Conclusion}
In this paper, we compared uniform and random sampling schemes for speech and music signals. We also used different reconstruction techniques for both sampling schemes. In case the speech signal is divided into frames, both objective performance metric criteria and subjective evaluations proposed that the uniform sampling with spline interpolation outperforms other methods. This is true due to the fact that speech and music signals are LP signals. However, this is not true if the signal has high frequency components or it is sparse. In case, the signal is not purely low-pass or sparse, one can use sub-band coding in which the low and high frequency components are sampled uniformly and randomly, respectively.

\bibliographystyle{IEEEtran}
\bibliography{Citations}

\end{document}